\documentclass[reprint,amsmath,amssymb,aps,prb,]{revtex4-2}

\usepackage{graphicx}
\usepackage{dcolumn}
\usepackage{bm}
\usepackage{color}

\begin{document}


\title{Engineering Ponderomotive Potential for Realizing $\pi$ and $\pi/2$ Bosonic Josephson Junctions}

\author{Jiadu Lin$^{1,2}$,  Qing-Dong Jiang$^{1,2,}$}
\email{qingdong.jiang@sjtu.edu.cn}
\affiliation{$^1$Tsung-Dao Lee Institute \& School of Physics and Astronomy, Shanghai Jiao Tong University, Pudong, Shanghai, 201210, China\\
$^2$Shanghai Branch, Hefei National Laboratory, Shanghai, 201315, China.
}

\begin{abstract}
We study the ponderomotive potential of a bosonic Josephson junction periodically modulated by a high-frequency electromagnetic field. Within the small population difference approximation, the ponderomotive drive induces the well-known Kapitza pendulum effect, stabilizing a $\pi$-phase mode. We discuss the parameter dependence of the dynamical transition from macroscopic quantum self-trapping to $\pi$-Josephson oscillations. Furthermore, we examine the situation where the small population difference approximation fails. In this case, an essential momentum-shortening effect emerges, leading to a stabilized $\pi/2$-phase mode under certain conditions. By mapping this to a classical pendulum scenario, we highlight the uniqueness and limitations of the $\pi/2$-phase mode in bosonic Josephson junctions.



\end{abstract}

\maketitle


\section{\label{sec:level1}Introduction}

The superconducting Josephson effect was first predicted in 1962 \cite{JOSEPHSON1962}, showing that the cooper pair can tunnel through a thin insulator between two superconductors. The tunneling supercurrent depends on the phase difference across the tunneling barrier. Over past decades, a variety of current-phase relations in Josephson junctions have been theoretically discussed and experimentally observed \cite{Golubov2004}. For example, under certain conditions, a Josephson junction could sustain a phase difference of $\pi$ between the two superconductors in its ground state, a configuration known as a ``$\pi$- Josephson junction" \cite{bulaevskii1977}. Other types of nontrivial Josephson junctions, including $\pi/2$-phase junction, junctions exhibiting $0-\pi$ transitions, and tilted topological junctions were later studied in systems with broken time-reversal symmetry \cite{Buzdin2005,Zyuzin2000,Yuan2023,Yang2018,PhysRevB.98.054508,PhysRevLett.124.197001,PhysRevLett.129.016801,Yerin2014}.

The Bose-Einstein condensates (BECs) offer a charge neutral platform to explore the Josephson effect, where neutral atoms replace electrons in the current. It was first proposed in two weakly coupled Bose-Einstein condensates in a double-well trap \cite{Javanainen1986}, known as the Bose-Josephson junction (BJJ) or the bosonic Josephson junction. A macroscopic number of bosons condensate in the two wells, and the weak coupling between them gives rise to the quantum tunneling phenomenon. Furthermore, interactions between bosons induce nonlinear effects, often described by the nonlinear Schr{\"o}dinger equation or Gross–Pitaevskii equation. The physics of a classical pendulum is often used as an analogy for understanding different dynamic modes in the BJJ, including the traditional Josephson tunneling, self-trapping and $\pi$-phase oscillations \cite{Smerzi1997,Zapata1998,Raghavan1999,Raghavan1999tran,shenoy2002tunneling}. In experiments, a single bosonic Josephson junction was first realized in the optical lattice \cite{Albiez2005}. Later the d.c. Josephson effects \cite{levy2007}, interatomic interactions \cite{Spagnolli2017}, phase-locking \cite{Pigneur2018,Pigneur2018analytical} and spin squeezing \cite{Zhang2024} were also investigated in experiments.

When some parameters are periodically modulated in the BJJ, different phenomena can arise \cite{Wang2006,Lin2023,Liu2024,Singh2024}. For example, by oscillating the energy bias between the two traps with the high frequency, phase-locking \cite{kohler2003chemical,Xie2008} and quasienergies \cite{Luo2008} have been studied. Moreover, research shows that the time-dependent nonlinear interactions may drive the system into chaos through period doubling bifurcations \cite{fei2007atomic}. The periodically driven Josephson coupling energy can be associated with chaos \cite{Abdullaev2000,Liu2021}, quantum metrology \cite{Liu2021}, quantum simulation \cite{Zhu2021,wang2024}, ring structures \cite{Lyu2020} and Landau-Zener tunneling \cite{Zhang2008,Cardoso2024}. And if the modulation originates from an external pulse field, the system resembles the nonlinear Rosen-Zener model \cite{Ye2008,Ye2008Ramsey}. Experimentally, Floquet engineering of a one-dimensional BJJ has been achieved in a tilted double-well potential \cite{Ji2022}.

There is a non-intuitive phenomenon resulting from the external modulation in classical mechanics, showing that an upside-down pendulum could be stabilized by a vertical high-frequency force. This physical phenomenon is known as the Kapitza effect \cite{Kapitsa1951}. Applying the Kapitza effect to the BJJ, the collective dynamics under a fast driving field with a noisy component was studied and a stabilized $\pi$-phase state was predicted \cite{Boukobza2010} using the master equation. However, in previous works, the effective stationary potential and its parameter dependence of this $\phi=\pi$ coherent state were not fully explored. In this paper, instead of using the master equation, we introduce the ponderomotive potential \cite{LANDAU197658,Aliev1992} to reveal the dynamics for a large time scale without noise. Ponderomotive potential, or ponderomotive force, appears widely in laser physics, describing the motion trend of charged particles in an inhomogeneous oscillating electromagnetic field \cite{Aliev1992,Rahav2003}. Now this notion has been extended to quantum many-body physics \cite{Sun2018,Wolff_2019,Rikhter2024,Sun2024dynamical,Sun2024Floquet} according to the similar physical picture that a system undergoes an inhomogeneous periodic driving field and exhibits distinct behaviors across different time scales. 

By using the ponderomotive potential, we investigate how the periodic modulation of Josephson coupling induces transitions between different dynamic modes. First, we apply periodic modulation to the BJJ and show the time-dependent equations of motion under the two-mode approximation \cite{Smerzi1997}, using the macroscopic wave function. For a small population difference, the analytically calculated effective potential reconfirms the energy minimum with a phase difference of $\pi$. We analyze the parameter dependence to understand how the dynamic modes are regulated by the driving frequency and amplitude, causing the dynamical phase transition. In addition to $\pi$-modes, $\pi/2$-modes are more intriguing because, in superconductor Josephson junctions, time-reversal symmetry breaking needs to be broken to form a $\pi/2$-junction. In our setup, the periodic driving field to the BJJ provides the necessary time-reversal symmetry breaking to form $\pi/2$-modes. We go deep into the unique momentum-shortening effect of the BJJ and reveal the possibility of stabilized $\pi/2$-phase dynamic modes, where the small population difference approximation is no longer available. Analytical results of relative phase behavior in the vicinity of $\pi/2$-phase region are derived under a driving field, and these findings are supported by numerical simulations. Finally, we reveal the dependency of the $\pi/2$ modes' lifetime on the time-averaged tunneling energy.

\section{A Bosonic Josephson Junction}
\begin{figure}[t]
\includegraphics[width=3in]{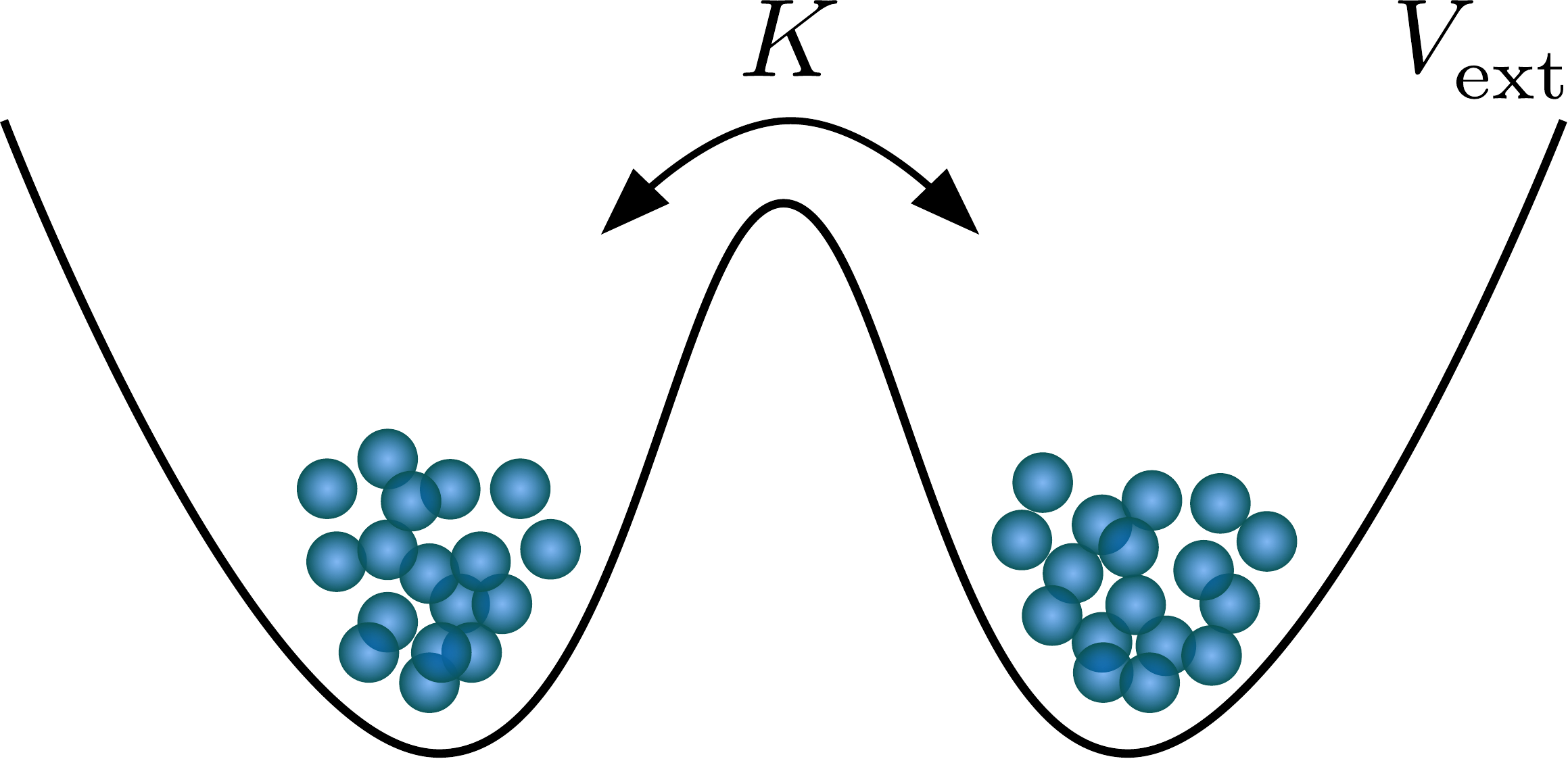}
\caption{\label{system} A bosonic Josephson junction. Cold atoms are constrained in the two traps by a double-well potential. The high barrier potential between two condensates leads to weak coupling as well as the tunneling effect. }
\end{figure}
As schematically shown in Fig.\,\ref{system}, a bosonic Josephson junction is essentially constructed by two weakly coupled BECs in a double-well trap in optical lattices. One could use a many-body wave function $\Psi(x, t)$ to characterize the quantum state of a macroscopic number of neutral ultracold atoms. The many-body wave function satisfies the Gross–Pitaevskii equation
\begin{equation}
\begin{aligned}
     i\hbar\frac{\partial \Psi(\vec{r},t) }{\partial t}=&-\frac{\hbar^{2}}{2m}\nabla^{2}\Psi(\vec{r},t)+\Big [V_{\mathrm{ext}}(\vec{r})\\
    &+g_{0}\left | \Psi(\vec{r},t)\right |^{2} \Big ]\Psi(\vec{r},t),
\end{aligned}
\label{GPE}
\end{equation}
where $g_0=4\pi \hbar a_s/m$ describes the interatomic potential reduced from scattering process, and $a_s$ is the s-wave scattering length of atoms. $V_{\mathrm{ext}}$ denotes the external double-well potential. As in Fig.\ref{system}, the atoms reside respectively in the two traps separated by the high barrier potential. Thus the well-known two-mode approximation is adopted, assuming the macroscopic wave function to be the superposition of two wave functions describing the condensate in each trap,
\begin{equation}
\begin{aligned}
\Psi(\vec{r},t)&=\sqrt{N_1(t)}e^{i\phi_1(t)}\varphi_1(\vec{r})+\sqrt{N_2(t)}e^{i\phi_2(t)}\varphi_2(\vec{r})\\
    &=\Psi_1(t)\varphi_1(\vec{r})+\Psi_2(t)\varphi_2(\vec{r})
\end{aligned}
\label{wavef}
\end{equation}   
with $\Psi_1(t)$ and $\Psi_2(t)$ being complex time-dependent amplitudes, $\varphi_1(\vec{r})$ and $\varphi_2(\vec{r})$ being single particle ground-state solutions in the two wells, respectively. $N_{1,2}(t)$ and $\phi_{1,2} (t)$ are the number of atoms and the phases of the condensate in each trap. It can be considered as a mean-field description. Substituting Eq.\eqref{wavef} into Eq.\eqref{GPE} , we obtain the coupled Schr{\"o}dinger equation
\begin{equation}
\left\{
             \begin{array}{lr}
             i\hbar \dot{\Psi}_1(t)=(E_1+U_1 N_1)\Psi_1(t)-K\Psi_2(t) ,\\
             i\hbar \dot{\Psi}_2(t)=(E_2+U_2 N_2)\Psi_2(t)-K\Psi_1(t) .
             \end{array}
\right.
\end{equation}
Here $E_{1,2}$ are the zero-point energies in each well. $U_1 N_1, U_2 N_2$ represent the total atomic self-interaction energies and $K$ describes the Josephson tunneling energy between the condensates. For $U _ { 1,2 }$ there are initially several terms containing wavefunction overlap between two traps, which could be neglected due to the weak coupling condition. Then the parameters are
\begin{subequations}
\begin{align}
E _ { 1,2 } &= \int \left [ \frac { \hbar ^ { 2 } } { 2 m } \left | \nabla \varphi _ { 1,2 }  \right | ^ { 2 } + \varphi _ { 1,2 } ^ { 2 }   V _ { \mathrm{ext} }   \right ] d \vec { r }  \label{E} ,\\
 U _ { 1,2 } &= g _ { 0 } \int \varphi _ { 1,2 }^ { 4 }   d \vec { r } \label{U} ,\\
K &= - \int \left [ \frac { \hbar ^ { 2 } } { 2 m } \left ( \nabla \varphi _ { 1 }   \cdot \nabla \varphi _ { 2 }   \right ) + \varphi _ { 1 }   \varphi _ { 2 }   V _ { \mathrm{ext} }  \right ] d \vec { r } \label{K}.
\end{align}
\end{subequations}
Integrating over spatial coordinates and rescaling, we derive the dimensionless equations of motion of the two variables $n=\frac{N_1-N_2}{N_1+N_2} , \theta=\phi_2-\phi_1$,
\begin{equation}
\left\{
             \begin{array}{lr}
             \dot{n}=-\sqrt{1-n^2}\sin{\phi},\\
             \dot{\phi}=\frac{n}{\sqrt{1-n^2}}\cos{\phi}+\Lambda n+\mathrm{\Delta} E . 
             \end{array}
\right.\label{eqn}
\end{equation}
Here $\mathrm{\Delta} E=(E_1-E_2)/2K+(U_1-U_2)(N_1+N_2)/4K$ is the total energy bias between two traps, $\Lambda=(U_1+U_2)(N_1+N_2)/4K$ is the interatomic energy, and the time is rescaled $\frac{2K}{\hbar}t \rightarrow t$. The exact solutions of the two differential equations are simply Jacobian elliptic functions. 

In the special situation where the double-well potential is symmetric and $\mathrm{\Delta} E=0$, the equations of motion give rise to three different dynamic modes: Josephson oscillations, self-trapping, and $\pi$-Josephson oscillations. Josephson oscillations denote the states that both $n$ and $\phi$ oscillate around zero, while being harmonic with small amplitude (plasma oscillations) and anharmonic with large amplitude. It is quite similar to the ac Josephson effect in superconductor Josephson junctions. However, in a certain region, the time-averaged population difference is nonzero, which indicates the number of bosons in one well is always larger than the number in another. Meanwhile, the phase difference increases all the time. Thus the oscillation modes are called self-trapping or running-phase modes. As for the $\pi$-Josephson oscillations, $n$ still oscillates around zero but the average phase difference is $\pi$. Obviously they have a similar property of $\pi$-junctions in superconductor Josephson junctions that the free energy is minimized for the phase difference $\phi=\pi$ state.

Using Eq.\eqref{eqn}, it's straightforward to derive a classical Hamiltonian
\begin{equation}
H=\frac{\Lambda}{2}n^2-\sqrt{1-n^2}\cos{\phi}+\Delta En
\end{equation}
with $n$ and $\phi$ to be the canonical variables satisfying $ \dot{n}=-\frac{\partial H}{\partial \phi},\dot{\phi}=\frac{\partial H}{\partial n}$ (the canonical equation).

The Hamiltonian resembles that of a pendulum system in classical mechanics, with angular displacement $\phi$, angular momentum $n$, moment of inertia $1/\Lambda$, and applied torque $\mathrm{\Delta} E$. This analogy provides an intuitive understanding of the three different modes. Josephson oscillations correspond to the downward-orientation pendulum. As the amplitude increases, the oscillation transitions from harmonic to anharmonic. If the pendulum has sufficient energy to get over the peak, then it can twirl around with nonzero angular momentum and increasing angular displacement, similar to the self-trapping modes. $\pi$-Josephson oscillations correspond to an upward-orientation pendulum, arising from the nonrigid momentum-shortening effect.

\section{Ponderomotive potential method}
We now consider generalizing the BJJ by applying a periodic modulation in the harmonic form. It has been shown that, with an off-resonant drive, the many-body equivalents of the ``inverted pendulum’’ state can be achieved in the BJJ, analogous to the classical Kapitza pendulum \cite{Boukobza2010}. Within the previous work, the method of time scale separation is discarded while considering a noisy component in the driving field, and the master equation is applied. Nevertheless, instead of invoking the master equation, we introduce the ponderomotive potential to investigate the time evolution of the BJJ in the absence of noise.

Considering an external potential that oscillates periodically in the form $V(\vec{r},t)=V_0(\vec{r})+V'(\vec{r})\cos{\Omega t}$, the parameters $K$ and $\Delta E$ become time-dependent. Here, we only focus on the oscillation of the tunneling energy $K$ and set the energy bias $\Delta E$ to zero by tuning both $V_0(\vec{r})$ and $V'(\vec{r})$ to be spatially symmetric. Experimentally, the oscillation can be realized by periodically modulating the laser intensity or performing an extra laser beam to change the barrier potential between the two traps. Taking into account the high-frequency limit, we set $\Omega$ to be much larger than the characteristic frequency of the system. The parameter $K(t)$ can then be written as
\begin{equation}
K(t)=K_0+K'\cos{\Omega t}
\end{equation}
with the oscillation amplitude $K'$ determined by $V'(\vec{r})$,
\begin{equation}
K'=\int\varphi_1(\vec{r}) \varphi_2(\vec{r}) V'(\vec{r}) d\vec{r}.
\end{equation}
Using the same method, we get the dimensionless equations of motion
\begin{equation}
\left\{
             \begin{array}{lr}
             \dot{n}=-\sqrt{1-n^2}\sin{\phi}(1-\lambda \cos{\omega t}),\\
             \dot{\phi}=\frac{n}{\sqrt{1-n^2}}\cos{\phi}(1-\lambda \cos{\omega t})+\Lambda n . 
             \end{array}
\right.\label{eom_k0}
\end{equation}
The dimensionless parameter $\lambda$ denotes the ratio of $K'$ to $K_0$, $\lambda=K'/K_0$. The time is still rescaled $\frac{2K}{\hbar}t \rightarrow t$ and the frequency also becomes dimensionless $\omega=\frac{\hbar}{2K}\Omega$. When the relative oscillation amplitude $\lambda$ is rather small, commonly used perturbation techniques can be applied \cite{Abdullaev2000,Xie2003}. However, here we consider a more general case for $\lambda$ and turn to the high-frequency limit to obtain a time-independent effective potential.

\subsection{$\pi$-phase modes}
\begin{figure}[b]
\includegraphics[width=3in]{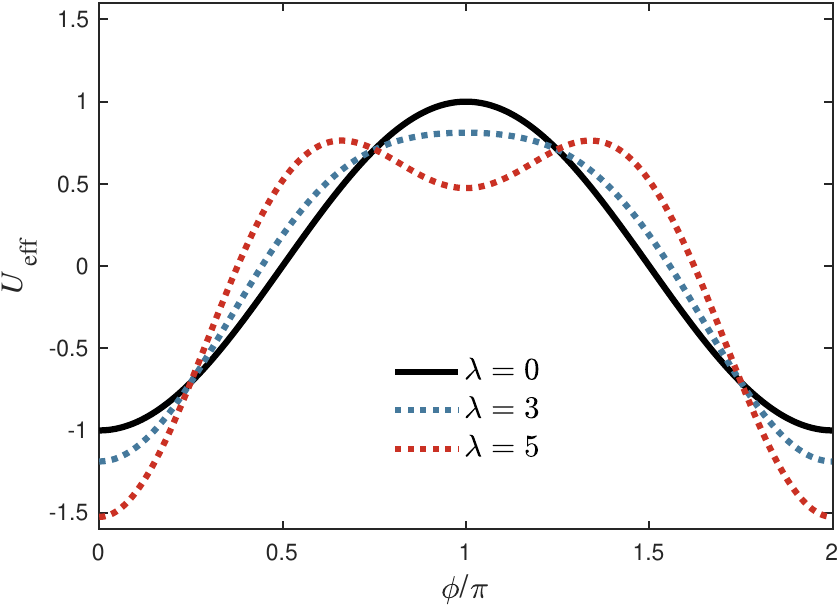}
\caption{\label{Ueff_fig} Lines with different colors show effective potential with respect to $\phi$ at different oscillation strength $\lambda$. The dimensionless oscillation frequency is $\omega=12$. The dimensionless interatomic energy is $\Lambda=11$.}
\end{figure}
The time-dependent terms complicate the equations, so here we apply a small $n$ approximation to study the small-amplitude oscillations around stationary states. 
Using a technique similar to that for analyzing $\pi$-Josephson oscillations \cite{Raghavan1999} in conventional BJJs, we drop higher-order terms of $n$ and derive the equation for $\phi$
\begin{equation}
\frac { 1 } { \Lambda } \ddot { \phi } = - \sin{\phi} + \lambda \sin{\phi}  \cos{\omega t} - \frac { \lambda ^ { 2 } } { \Lambda } \sin{\phi}  \cos{\phi }  \cos^2{ \omega t } .
\end{equation}
Since the dimensionless frequency $\omega$ is quite high, we separate the equation to two distinct time scales. Rapid oscillation terms are cast away as nonessential details. Thus we can give a time-independent effective potential as a function of $\phi$, which originates from the rapid oscillations,
\begin{equation}
U_{\rm eff}=-\cos{\phi}-\frac{\lambda^2}{8}\left(\frac{\Lambda}{\omega^2}+\frac{1}{\Lambda}\right)\cos{2\phi}-\frac{\lambda^4}{128\Lambda\omega^2}\cos{4\phi} \label{Ueff}.
\end{equation}
The derivation is the same as the classical one, depicting the motion in a rapidly oscillating field \cite{LANDAU197658,Rahav2003}. The effective potential here is known as the {\it ponderomotive potential}.

Because of the ponderomotive potential, $\phi=\pi$ can potentially serve as a stable minimum if the second derivative $\left.\frac{d^2 U_{\rm eff}}{d\phi^2}\right|_{\phi=\pi}$ is greater than $0$, as shown in Fig.\ref{Ueff_fig}. In that case, the system can be restricted to oscillate around $\phi=\pi$ with appropriate conditions, which can be recognized as $\pi$-Josephson oscillations. This suggests that, with the same initial condition and parameter, the applied high-frequency periodic modulation can induce a dynamic modes transition from self-trapping to $\pi$-Josephson oscillations.

However, it is worth noticing that the oscillation parameters $\lambda$ and $\omega$ are required to meet several conditions: (a) $\omega \gg 1$, indicating the limit of high frequency. (b) $\frac{\omega^2}{\lambda} \gg \Lambda$, meaning that the fast part of the motion should have a small oscillation amplitude compared with the slow one. The first two conditions guarantee that the derivation of the ponderomotive potential is valid. (c) $\lambda > \sqrt{\frac{2\Lambda}{\Lambda^2+\omega^2}}\omega$, which is indispensable to form a valley around $\phi=\pi$ within the poderomotive potential. With a fixed large frequency $\omega$, we can study the dynamical phase transition with respect to $\lambda$. When $\lambda$ is not large enough, the system remains in self-trapping modes since the ponderomotive potential is not strong enough to constrain it. By gradually increasing the driving strength, it is straightforward to see the transition from self-trapping modes to $\pi$-Josephson oscillations with the same initial condition. For the regular BJJ, in order to have $\pi$-Josephson oscillations, the self-interaction needs to be weak enough ($\Lambda<1$). For ponderomotive $\pi$-BJJs, however, such dynamic modes are feasible for arbitrary $\Lambda$ as long as $\lambda$ and $\omega$ meet the conditions. 

\begin{figure}[htb]
\includegraphics[width=3in]{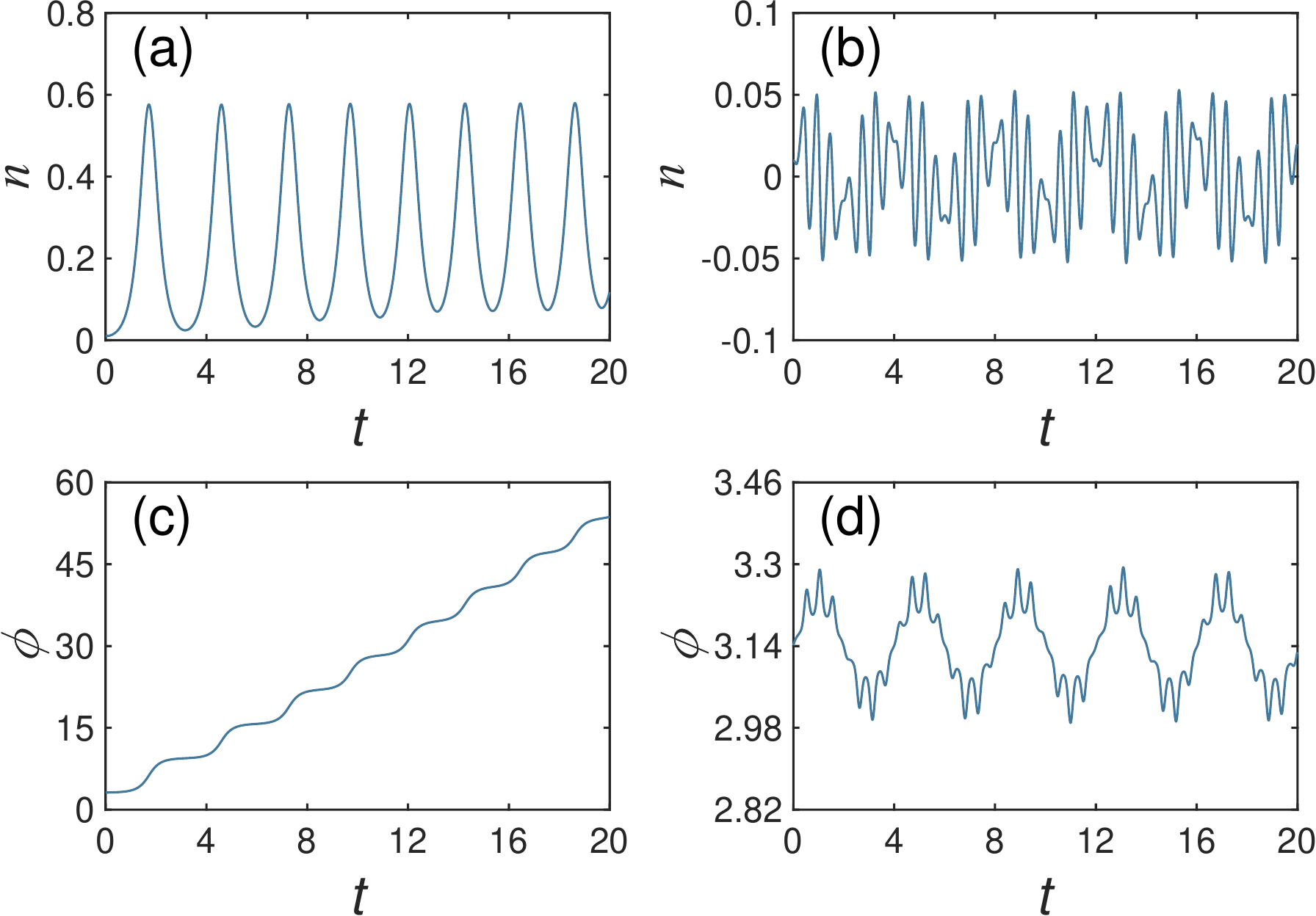}
\caption{\label{mode} Particle number difference $n$ and phase difference $\phi$ as a function of dimensionless time. The left two figures (a) and (c) show the numerical result of equations of motion without the driving field. Meanwhile, the right two figures (b) and (d) show the motion with the driving field. For both the two situations we have $\Lambda=11$ and the same initial conditions: $n(0)=0.01, \phi (0)=\pi$. And the parameters of the driving field are $\lambda = 6, \omega=12$. }
\end{figure}

In order to show the consistency, we numerically simulate the evolution of Eq. \eqref{eom_k0}, as shown in Fig.\,\ref{mode}. The result reveals the fact that the periodic modulation of the BJJ can lead to the transition of dynamic modes from self-trapping to $\pi$-Josephson oscillations in the same initial condition, in agreement with the analytical one. And we can see the rapid oscillations of variable $n$ and $\phi$ in Fig. \ref{mode}(b) originate from the high-frequency term. It can be cast away, which will not affect the tendency over periods, consistent with the time scale separation method we use to derive ponderomotive potential.

\subsection{Analogy to pendulum}
With the time-dependent equations of motion Eq.\eqref{eom_k0}, we can still write down the classical Hamiltonian
\begin{equation}
H=\frac{\Lambda}{2}n^2-\sqrt{1-n^2}\cos{\phi}(1-\lambda \cos{\omega t})\label{Ht}.
\end{equation}
Then the dynamic modes can still analogously be described using a periodically driven pendulum whose suspension point vertically oscillates. Choosing the suspension point frame as the moving coordinate frame, there's an additional oscillating term originating from the vertical acceleration, similar to Eq.\eqref{Ht}. The most interesting thing is, if the applied oscillation frequency is sufficiently high, the pendulum can be stabilized at the upward position---forming an inverted pendulum. With the method of time scale separation or Floquet discriminant, this phenomenon can be analytically explained, which was first pointed out by Kapitza in 1951 \cite{Kapitsa1951}. Here, we see $\pi$-Josephson oscillation discussed above is a realization of Kapitza pendulum in the BJJ. 

\subsection{Zero-mean Josephson coupling energy}
When considering the special case that $K_0=0$, there's only an oscillation term with the time-averaged tunneling energy to be 0, but we can still see the fluctuation of the tunneling energy
\begin{equation}
\Delta K=\sqrt{\overline{K^2(t)}-\overline{K(t)}^2}=\frac{K'}{\sqrt{2}}.
\end{equation}
Here $\overline{K(t)}$ denotes the average of $K(t)$ over one period.

Such a zero-mean tunneling energy has been introduced in the BJJ for the sake of chaos behavior \cite{Liu2021}. For superconducting Josephson junctions, the coupling between two superconductors can be readily modulated by tuning the property of the material in between, from insulators to ferromagnetic materials. Likewise, it is proposed that with a mixture of BECs, one component of which resides in the center and acts as a medium when tunneling happens \cite{Maraj2017,Niu2019}, the tunneling parameter is precisely tunable. 


Then the parameter $\lambda=K'/K_0$ diverges to infinity and we should rewrite the formula of equations of motion
\begin{equation}
\left\{
             \begin{array}{lr}
             \dot{n}=\sqrt{1-n^2}\sin{\phi}\cos{\omega' t},\\
             \dot{\phi}=-\frac{n}{\sqrt{1-n^2}}\cos{\phi} \cos{\omega' t}+\Lambda' n , 
             \end{array}
\right.\label{eom0}
\end{equation}
 and the ponderomotive potential
\begin{equation}
U_{\rm eff}=-\frac{1}{8}\left(\frac{ \Lambda'}{\omega'^2}+\frac{1}{ \Lambda'}\right)\cos{2\phi}-\frac{1}{128 \Lambda'\omega'^2}\cos{4\phi} \label{Ueff'}.
\end{equation}
Here time is rescaled $\frac{2K'}{\hbar}t \rightarrow t$, dimensionless frequency $\omega'=\frac{\hbar}{2K'}\Omega$, and self-interaction energy $\Lambda'=(U_1+U_2)(N_1+N_2)/4K'$.

From Eq.\eqref{Ueff} to Eq.\eqref{Ueff'}, the $\cos\phi$ term disappears and $\phi=\pi$ turns out to be a minimum regardless of the oscillation amplitude $K'$ . At the same time, it is a global minimum instead of a local one in the previous situation, which suggests a wider initial condition regime for $\pi$-Josephson oscillations. Likewise, we can turn to the pendulum picture to understand: $K_0=0$ indicates there's no gravity applied on the pendulum, and the static field $\cos\phi$ term vanishes. Imagining the case where the pendulum is laid flat and is driven in one direction, the ponderomotive force will dominate and form two stable points in that direction.

\section{$\pi/2-$phase mode}

In the previous situation, we consider the small $n$ limit to derive the ponderomotive potential (Eq. \eqref{Ueff}) with respect to the phase difference $\phi$, which results in the $\pi$-phase dynamic modes. Nevertheless, within such an approximation, it is not sufficient to reveal all the dynamic information for the system. We can show that the time-independent effective potential will be slightly deformed in the vicinity of $\phi=\pi/2$ if we take the variation of $n$ into account, which means that the so-called momentum-shortening effect is no longer negligible. Then with properly selected parameters, an effective potential minimum shows up at $\phi=\pi/2$, and, therefore, a $\pi/2$-phase mode appears.

Following discussion in the $K_0=0$ case, we set the new variable $\theta=\phi-\pi/2$ in order to investigate the equations of motion Eq. \eqref{eom_k0} around $\phi=\pi/2$,
\begin{equation}
\left\{
             \begin{array}{lr}
             \dot{n}=\sqrt{1-n^2}\cos{\theta}\cos{\omega' t},\\
             \dot{\theta}=\frac{n}{\sqrt{1-n^2}}\sin{\theta} \cos{\omega' t}+\Lambda' n .
             \end{array}
\right.\label{eom_th}
\end{equation}

To analyze the dynamic modes, we linearize the equation above with the ansatz $\theta \ll 1$. Later, we will check if the result is consistent with our ansatz. Thus for the first equation $n$ and $\theta$ are decoupled and a closed-form expression of $n$ can be immediately calculated,
\begin{equation}
    n(t)=\sin{\left(\frac{1 }{\omega'}\sin{\omega' t}+C_1\right)}.
\label{n}
\end{equation}

The constant $C_1$ here is determined by the initial condition $n(0)$. For convenience we assume $n(0)=0$, which is reasonable that the two traps acquire an equal number of particles initially, in conditions that symmetric external potential $\Delta E=0$ and repulsive interaction $\Lambda' > 0$.

The expression Eq. \eqref{n} can be rewritten with the Jacobi-Anger expansion and the coefficients are Bessel functions of the first kind (see Appendix A). With a sufficiently high frequency, the characteristics of Bessel function imply that the first order term in the expansion dominates,
\begin{equation}
    \begin{aligned}
        n(t)=&\sin{(\omega'^{-1}\sin{\omega' t})}\\
        \simeq &2J_1(\omega'^{-1})\sin{\omega' t}.
    \end{aligned}\label{n_bessel}
\end{equation}
Here $J_n$ represents the $n$th order of Bessel function. Substituting this expression into the equation of $\dot{\theta}$, we get
\begin{equation}
\dot{\theta}=\frac{J_1(\omega'^{-1})}{J_0(\omega'^{-1})}\sin(2\omega' t) \cdot \theta+2 \Lambda' J_1(\omega'^{-1})\sin{\omega' t}.
\label{theta_dot}
\end{equation}
The solution of such a linear first order differential equation is given by the method of constant variation,
\begin{equation}
    \begin{aligned}
        \theta(t)=&-\Lambda' \sqrt{\frac{\pi J_0 J_1}{\omega'}}\left[\mathrm{erf}\left(\frac{J_1}{J_0 \omega'}\cos{\omega' t}\right)-C_2\right]\\
        &\cdot \mathrm{exp}\left \{\frac{J_1}{2J_0 \omega'}(\cos{2\omega' t}+1) \right\}.
    \end{aligned}
\label{theta}
\end{equation}
Here $\mathrm{erf}(x)$ represents the error function of $x$. This analytical result indicates that $\theta(t)$ is periodic and oscillates around 0. However, the aforementioned expression on the one hand reveals too much redundant details we are not concerned about. For instance, we do not pay much attention to the fast ``micromotion" within a period. Instead, the long-timescale dynamics is the essential part in a realistic situation. On the other hand, it is not so intuitive and the physical picture needs to be clarified. 
\begin{figure}[b]
\includegraphics[width=3in]{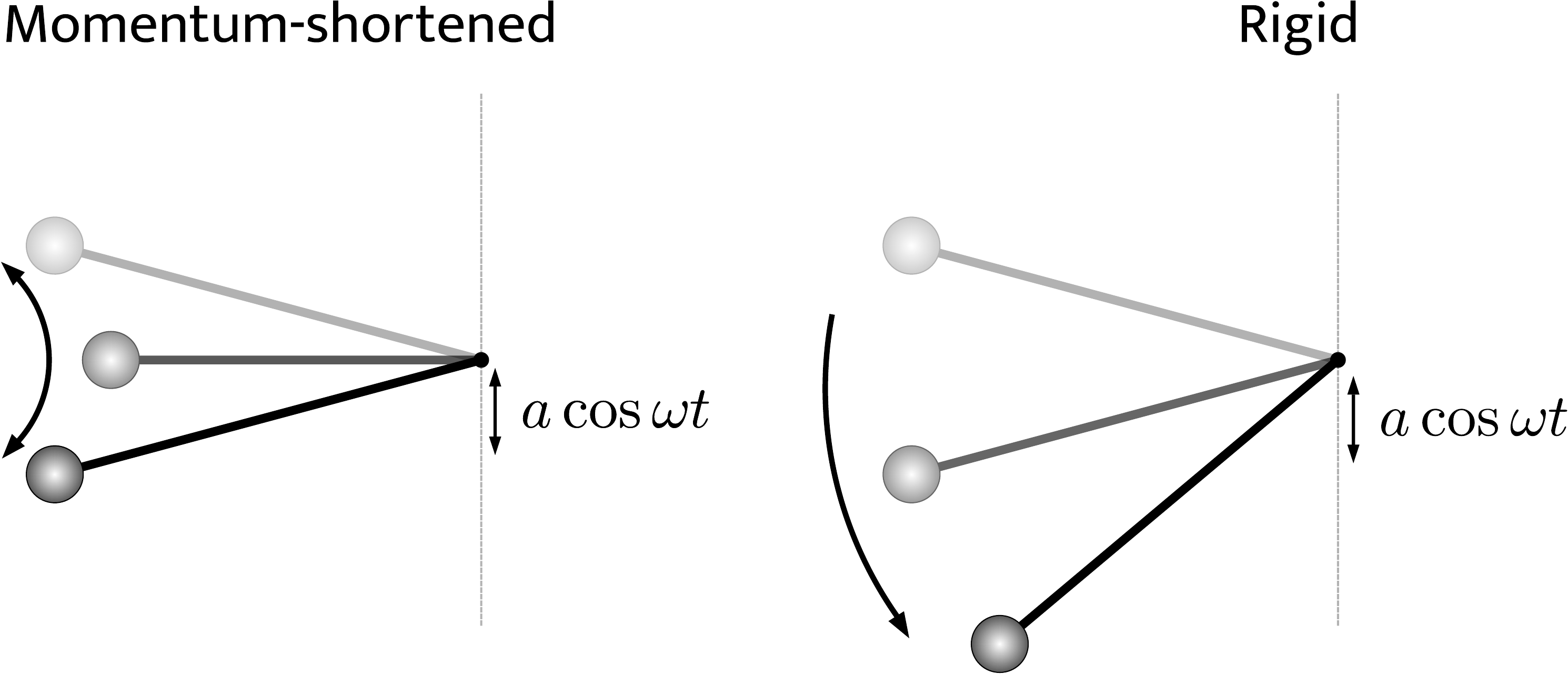}

\caption{\label{pendulum}The illustration of two different types of pendulum with the same vertical driving force. The left figure shows a momentum-shortened pendulum vibrates around $\pi/2$ point in a counter-intuitive way: the length of the rod shrinks and recovers according to its motion, which stabilizes the $\pi/2$ modes. While the classical pendulum with a rigid rod will just go through the $\pi/2$ point, as shown in the right figure.}
\end{figure}

From this point of view, with Eq. \eqref{theta_dot} we can derive the ponderomotive potential through the similar approach in section III,
\begin{equation}
 U'_{\mathrm{eff}}(\theta)=\left[\frac{J_1(\omega'^{-1})}{J_0(\omega'^{-1})}\right]^4\frac{1}{256\Lambda' \omega'^2}\theta^2.
\label{U_theta}
\end{equation}
Then we can recognize that compared with the original effective potential Eq. \eqref{Ueff'}, the result shows a potential minimum around $\theta=0$, or equivalently $\phi=\pi/2$, which stabilizes the $\pi/2$-phase modes. 

Looking back to the small $\theta$ ansatz, obviously it is satisfied with these two expressions Eq. \eqref{theta} and Eq. \eqref{U_theta}, showing that it is reasonable. However, the ansatz also implies that only in a small region around $\theta=0$ are these results valid. 

\subsection{Analogy to pendulum}

Still, we want to interpret this phenomenon with the analogy to a pendulum in the context of classical mechanics. As mentioned in the previous section, when we map it to a pendulum, the condition $K_0=0$ indicates that no gravity exists, rendering a pendulum laid flat. But the rotational symmetry is still broken by the periodic shaking. Then here the $\theta=0$ , i.e., $\phi=\pi/2$, direction refers to the orientation perpendicular to the periodic shaking. From the discussion of $\pi$-state we show that the high-frequency modulation will generate a stabilized state in the oscillation direction. But in the absence of gravity, the momentum-shortening effect will lead to an intriguing perpendicularly stabilized state, shown as the $\pi/2$-state. However, evidently this $\pi/2$-state won't appear for a rigid pendulum without the momentum-shortening effect, as illustrated in Fig.\ref{pendulum}. Instead it is a unique phenomenon in BJJs or, more generally, weakly coupled two-mode BECs. In that case, such a phenomenon could be generalized to other systems with Josephson-type effects \cite{Leggett2001}, for instance, the two weakly coupled superfluid reservoirs $^3$He and the internal Josephson effect between different hyperfine states.

\subsection{Discussion of results}

\begin{figure}[t]
\includegraphics[width=3in]{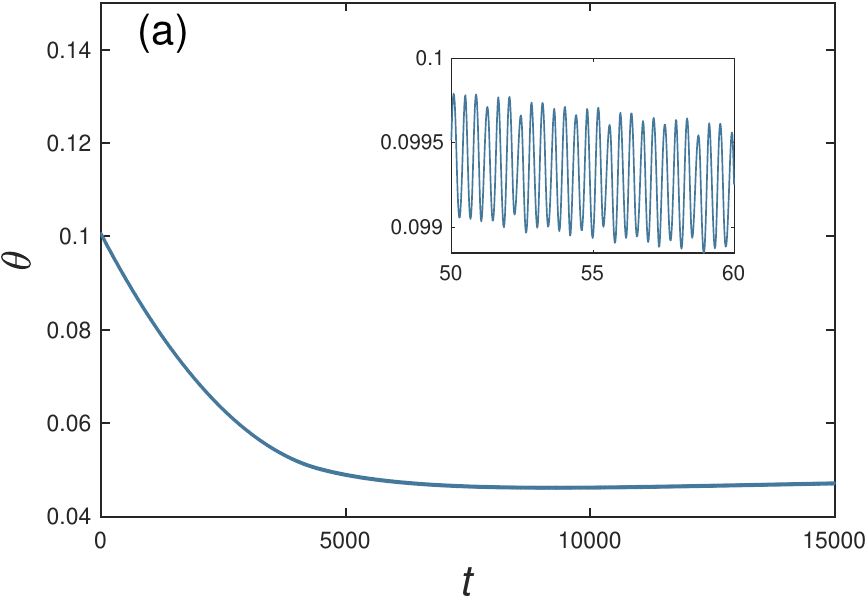}

\includegraphics[width=3in]{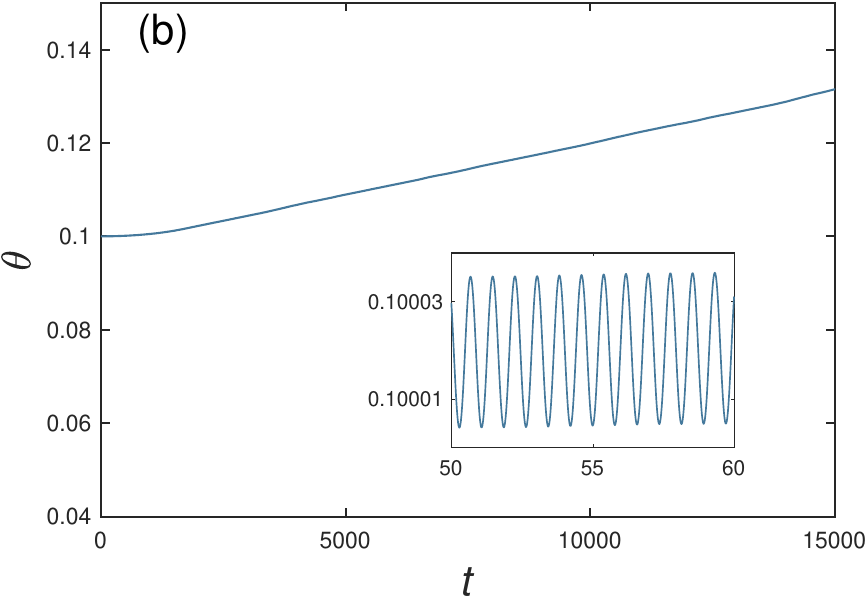}
\caption{\label{compare_cl}With high-frequency modulation and the condition $K_0=0$, numerical results of equations of motion corresponding to the BJJ (a) and classical rigid pendulum (b) are presented. Initial conditions are $n(0)=0, \theta (0)=0.1$ for both (a) and (b), parameters are $\Lambda' =1\times 10^{-3}, \omega'=8$ for both (a) and (b). In the context of classical pendulum, $1/\Lambda'$ is the moment of inertia of the rigid rod. The rapid dynamic behaviors are displayed in detail.}
\end{figure}

To verify our analytical results, we numerically solve Eq. \eqref{eom_th} as well, see Fig.\,\ref{compare_cl}. In contrast to an increasing deviation of $\theta=0$ in the regular rigid pendulum case, the numerical result of Eq. \eqref{eom_th} converges to $\theta=0$ and exhibits a stable $\pi/2$-phase mode over long time evolution,  in agreement with our analytical calculation. Although obtaining an explicit analytical expression for the initial value range in which $\pi/2$-phase modes appear is challenging, numerical simulation is feasible. For example, with the parameters in Fig.\,\ref{compare_cl}, to realize the $\pi/2$-phase mode, the upper bound of $\theta(0)$ is 0.152 and the lower bound is -0.148. It's natural that the region is not symmetric for the $\theta \rightarrow -\theta$ transformation, since we have artificially fixed the initial orientation of the driving force. That means, the universal form of oscillations should be $\cos(\omega' t+\phi_0)$, but we simply consider $\phi_0=0$. In the high-frequency case, such a preference will only lead to a slight difference in the absolute values of the upper and lower bounds, as we have shown.

For the sake of clarifying how the parameters play a role in the $\pi/2$-phase modes, we could dissect the two expressions Eq. \eqref{theta} and Eq. \eqref{U_theta}. For the modulation frequency $\omega'$, it must be large due to the high frequency approximation we apply as well as the employment of effective potential. Nonetheless, in Eq. \eqref{U_theta} combining the property of the Bessel function, we know that the strength of the effective potential is proportional to $\omega'^{-6}$, with the increasing frequency the range of the $\pi/2$-phase mode will be suppressed. Thus an intermediate-high $\omega'$ is appropriate to make the $\pi/2$-phase mode distinct. In a similar way, the effective potential necessitates a small $\Lambda'$ according to the effective potential, meanwhile, Eq. \eqref{theta} also agrees with such a limit. Because large $\Lambda'$ contributes to a large amplitude oscillation, conflicting with the ansatz $\theta \ll 1$. In experiments, the strength of self-interaction can be tuned in a large range using the Feshbach resonance technique; then achieving the parameter region of $\Lambda'$ is practical.

\begin{figure}[t]
\includegraphics[width=3in]{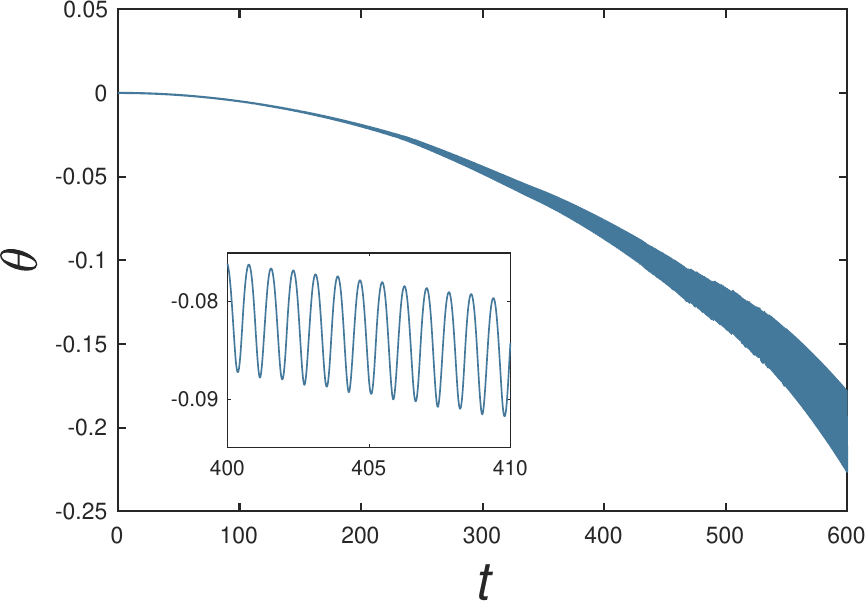}
\caption{\label{gammaterm}With high-frequency modulation and the condition $K_0\neq 0$, numerical result of equations of motion corresponding to the BJJ is presented. Initial conditions are set as $n(0)=0, \theta (0)=0$ , parameters are set as $\lambda =1\times 10^{-3}, \omega'=8,\gamma=1\times 10^{-3}$.}
\end{figure}

Another noteworthy scenario arises when the time-averaged tunneling energy has a finite value, implying $K_0 \neq 0$. Using a similar approach as in Eq. \eqref{theta}, the long-time asymptotic behavior can be derived. Defining $K_0/K'=\gamma$, when $t \rightarrow \infty$, we can know $\theta(t) \sim \mathrm{exp}\{\gamma^2 t^2/2\}$. This implies that a $\pi/2$-phase state can not sustain, as the deviation from zero grows over time, thereby violating the small $\theta$ approximation. This conclusion is supported by numerical simulation in Fig.\,\ref{gammaterm}. For another perspective, if $\gamma$ is zero the asymptotic term goes back to one, consistent with the previous results. For a small $\gamma$ the expression indicates that when $t \ll 1/\gamma$, the phase difference remains close to the $\pi/2$-phase region. In other words, the $\pi/2$-phase state has a finite lifetime on the order of $1/\gamma$, extending to infinity as $\gamma$ approaches zero.

\section{Conclusion}

We demonstrated the use of the ponderomotive potential method to clarify the relative phase dynamics in a BJJ with fast periodic modulation. Using time-scale separation, we quantitatively derived the effective potential governing the slow dynamics within the small population difference approximation. By tuning the driving field, we further explored the parameter dependence, revealing the region of $\pi$-Josephson oscillations and enhancing the understanding of the Kapitza effect in quantum systems. Additionally, we predict that when the time-averaged tunneling energy vanishes, $\pi/2$-phase modes emerge due to a unique momentum-shortening effect in BJJs. We emphasize that our analytical results, supported by numerical simulations, highlight a distinct difference between BJJs and classical rigid pendulums under high-frequency driving fields.

Finally, we investigate the case of non-zero time-averaged tunneling energy and show that even a small time-averaged tunneling energy disrupts the $\pi/2$-phase modes over long times. Through the ponderomotive method, we gained insights into the system’s relative phase dynamics, which could benefit future developments in matter-wave interferometry and quantum simulation.

Looking forward, rich phenomena could arise in the BJJ when subjected to high-frequency periodic modulation. However, the system's intrinsic nonlinearity complicates the analysis of its dynamics. For large oscillation amplitudes, theoretical methods become especially limited as they enter the non-perturbative regime. Questions remain regarding the full phase space behavior and the critical parameter values associated with dynamical mode transitions. Additionally, novel many-body states, such as the $\varphi$-junction \cite{Buzdin2003}, could be further explored by introducing asymmetry into the BJJ. Therefore, dedicated investigations into these aspects are required in the future.

\begin{acknowledgments}
We thank Augusto Smerzi, Zhiyuan Sun, Chao Gao and Botao Wang for helpful discussions and comments. This work was supported by National Natural Science Foundation of China (NSFC) under Grant No. 23Z031504628, Pujiang Talent Program 21PJ1405400, Jiaoda 2030 program WH510363001-1, the Innovation Program for Quantum Science and Technology Grant No. 2021ZD0301900, and the National Natural Science Foundation of China (NSFC) under Grant No. 12374332.

\end{acknowledgments}

\appendix

\section{Bessel functions of the first kind}

Bessel functions of the first kind, denoted as $J_n(x)$, are solutions of Bessel's differential equation. It is possible to define the function by its series expansion around $x = 0$, which can be found by applying the Frobenius method to Bessel's equation,
\begin{equation}
    J_\alpha(x)=\sum_{m=0}^{\infty} \frac{(-1)^m}{m ! \Gamma(m+\alpha+1)}\left(\frac{x}{2}\right)^{2 m+\alpha}.
    \label{bessel}
\end{equation}
Bessel’s equation emerges when considering the Helmholtz equation and Laplace’s equation in cylindrical or spherical coordinates with the method of separation of variables. Here we take advantage of the generating function of the Bessel function
\begin{equation}
    e^{\frac{x}{2}(t-\frac{1}{t})}=\sum_{n=-\infty}^{+\infty}J_n(x)t^n.
\end{equation}
The formula above plays an important role in the theory of Bessel functions for the series expansion. By setting $t=e^{i\theta}$, we can derive the following equations,
\begin{equation}
\left\{
             \begin{array}{lr}
             \sin{(x\sin{\theta})}=\sum_{n=-\infty}^{+\infty} J_n(x) \sin{n\theta},\\
             \cos{(x\sin{\theta})}=\sum_{n=-\infty}^{+\infty} J_n(x) \cos{n\theta}.
             \end{array}
\right.\label{eqnt}
\end{equation}
For integer order $n$, a special property can be used to make the expression more succinct, given by
\begin{equation}
    J_{-n}(x)=(-1)^n J_n(x).
\end{equation}
Now we turn to the calculation of particle number difference in $\pi/2$-phase modes. Invoking these mathematical techniques, we can rigorously rewrite Eq.\ref{n_bessel} in series expansion form
\begin{equation}
n(t)=\sin{\left(\frac{1}{\omega'}\sin{\omega' t}\right)}=\sum_{n=0}^{\infty} 2J_{2n+1}\left (\frac{1}{\omega'}\right ) \sin{(2n+1)\omega' t}.
\end{equation}
Furthermore, Eq.\ref{bessel} implies when $x\ll 1$, we can remain the lowest order of $x$ in $J_n(x)$ and get the approximate expression
\begin{equation}
    J_n(x) \sim \frac{1}{ \Gamma (n+1)}\left(\frac{x}{2}\right)^n.
\end{equation}
We can instantaneously figure out that a higher order term is much smaller in this limit,
\begin{equation}
    \frac{J_{n+1}(x)}{J_{n}(x)}\sim \frac{1}{(n+2)}\frac{x}{2} \ll 1.
\end{equation}
Thus, if the periodic driving field is fast enough and the oscillation frequency is sufficiently high, on the one hand, the coefficient of higher order sinusoidal terms should be small in contrast to the leading order; on the other hand, these higher order terms correspond to higher frequencies, which would hardly contribute to slow motion in a large time scale we are concerned about.

Combining all the arguments, the expression can be simplified to a neat sinusoidal form
\begin{equation}
    \begin{aligned}
        n(t)=&\sin{(\omega'^{-1}\sin{\omega' t})}\\
        \simeq &2J_1(\omega'^{-1})\sin{\omega' t}.
    \end{aligned}
\end{equation}

The similar employment of Bessel functions arises in Floquet systems \cite{Eckardt2017,Wang2018}. For example in a tight-binding lattice, the tunneling parameter would be changed to an effective one by a factor of the Bessel function with a strong time-periodic forcing, which can lead to dynamic localization \cite{Dunlap1986}.

\bibliography{main}

\end{document}